\documentclass[sigchi-a]{acmart}

\def\BibTeX{{\rm B\kern-.05em{\sc i\kern-.025em b}\kern-.08emT\kern-.1667em\lower.7ex\hbox{E}\kern-.125emX}}
    
\copyrightyear{2019}
\acmYear{2019}
\setcopyright{acmlicensed}
\acmConference[CHI '19 Extended Abstracts]{Conversational Agents: Acting on the Wave of Research and Development}{May 04--09}{Glasgow, UK}
\acmBooktitle{Conversational Agents: Acting on the Wave of Research and Development, May 04--09, 2019, Woodstock, NY}

\usepackage{graphicx,caption}

\begin{document}

%
\title{Designing Style Matching Conversational Agents}

%

\author{Deepali Aneja}
\email{deepalia@cs.washington.edu}
\affiliation{%
 \institution{University of Washington}
 \city{Seattle}
 \state{Washington}
 \country{USA}}

\author{Rens Hoegen}
\email{rhoegen@ict.usc.edu}
\affiliation{%
  \institution{University of Southern California}
  \city{Los Angeles}
  \state{California}
  \country{USA}}
  
\author{Daniel McDuff}
\email{damcduff@microsoft.com}
\affiliation{%
 \institution{Microsoft Research}
 \city{Seattle}
 \state{Washington}
 \country{USA}}
 
 \author{Mary Czerwinski}
 \email{marycz@microsoft.com}
\affiliation{%
 \institution{Microsoft Research}
 \city{Seattle}
 \state{Washington}
 \country{USA}}

%

%
\begin{abstract}
Advances in machine intelligence have enabled conversational interfaces that have the potential to radically change the way humans interact with machines. However, even with the progress in the abilities of these agents, there remain critical gaps in their capacity for natural interactions. One limitation is that the agents are often monotonic in behavior and do not adapt to their partner. We built two end-to-end conversational agents: a voice-based agent that can engage in naturalistic, multi-turn dialogue and align with the interlocutor's conversational style, and a 2nd, expressive, embodied conversational agent (ECA) that can recognize human behavior during open-ended conversations and automatically align its responses to the visual and conversational style of the other party. The embodied conversational agent leverages multimodal inputs to produce rich and perceptually valid vocal and facial responses (e.g., lip syncing and expressions) during the conversation. Based on empirical results from a set of user studies, we highlight several significant challenges in building such systems and provide design guidelines for multi-turn dialogue interactions using style adaptation for future research. 
\end{abstract}

%
%


%
\keywords{Embodied Conversational agents, conversational style, visual style, social behavior,  emotional expressions, social dialogue, multi-modality}

%

%
\maketitle

\section{Introduction}

A long-term goal of artificial intelligence has been to develop embodied and voice-based computer systems that are capable of natural language interactions. A host of science-fiction books and films have imagined how human-computer co-operation would be facilitated by such technology. This goal requires more than purely textual analysis of conversations. For example, the ``way" that people speak (e.g., prosody, pacing, and pauses) contains rich context for the interpretation of ``what" they say.  

Style matching is the phenomenon of a speaker adopting the behaviors or traits of their interlocutor.  This can occur through word choice, as in lexical entrainment as observed in human-human conversations~\cite{scissors2008linguistic,scissors2009cmc} and human-agent~\cite{shamekhi2016exploratory} interactions. It can also occur in non-verbal behaviors, such as with prosodic elements of speech~\cite{thomas2018style}, facial expressions and head gestures~\cite{cassell2000embodied} and other embodied forms~\cite{sejima2011virtual}. Again, non-verbal matching has been observed to affect human-human interactions~\cite{thomas2018style}, human-virtual agent interactions~\cite{levitan2016implementing,cassell2000embodied} and human-robot interactions~\cite{ono2001model}. It has also been shown that rapport and information disclosure is higher with assistants that can engage in social dialogue and respond appropriately ~\cite{bickmore2000weather,bickmore2001relational}

In our design for conversational agents, we implemented conversational and visual style matching. In terms of understanding human-to-human conversational style matching, Deborah Tannen's theory on conversational style ~\cite{tannen87style,tannen05style} is the most widely used. Tannen categorizes conversational style into high consideration (HC) and high involvement (HI). The HC style emphasizes consideration, independence and is characterized by slower and hesitant speech, longer pauses, and use of moderate paralinguistics. The HI style emphasizes interpersonal involvement, interest, understanding and is characterized by faster and louder speech with shorter pauses between conversational turns. 

The contributions of this work are to 1) present the challenges in building a voice-based and embodied conversational agent that is capable of extended (over 5-minute) conversations with a user, 2) discuss how these types of agents are understood and experienced by users, and 3) provide design guidelines for conversational systems with conversational and visual style matching.

\section{Style Matching}
In our work, the ECA performs visual and conversational style matching based on the interaction style of the user during the conversation, and the voice-based agent adapts to only the conversational style. The input data is collected by the audio and video pipelines (for the ECA) and by audio only (for the voice-based agent). We then analyze the data for specific style variables. These style variables are computed in real-time, while the user is interacting with the agent. The agent's response adapts to the style of the user by learning from the different style variables. The ECA learns and expresses through emotional expressions, head movement, text sentiment, and lip sync intensity (on the visual side) and learns the conversational style from the content of the participant's utterances like the use of pronouns, word repetition, utterance length, prosodic qualities, like speech rate, pitch, and loudness (on the conversational side). The voice-based agent performs the style adaptation only on the speaker's conversational style.

\section{User Studies}
We conducted two independent user studies (N=30 in both) in which participants talked with the voice-based agent and the ECA for 15 to 20 minutes, resulting in almost 10 hours of natural interaction data in each scenario. Both of the studies used between subjects design. In the voice-based agent user study, participants either interacted with an agent using conversational style matching (experimental), or one that did not (control). We conducted an independent user study with an ECA to investigate the combined effect of conversational and visual style matching on users' perceptions of the system comparing two interaction conditions - participants interacting with the ECA using visual and conversational style matching (experimental) and others interacting with the ECA with no style matching (control). 

We defined a set of day-to-day tasks (scenarios) that the participants would discuss with the agent to reach 15 minutes of interaction time with the agent. Each task involved a scenario that the participant was to discuss with the agent. The tasks centered around planning a lunch date, talking about personal life (where they were from, their likes and dislikes and their job), discussing a vacation to London, organizing a party and planning to go to a movie.

\begin{figure}[t!]
\vskip 0.2in
\captionsetup{width=\linewidth}
\centerline{\includegraphics[width=\columnwidth]{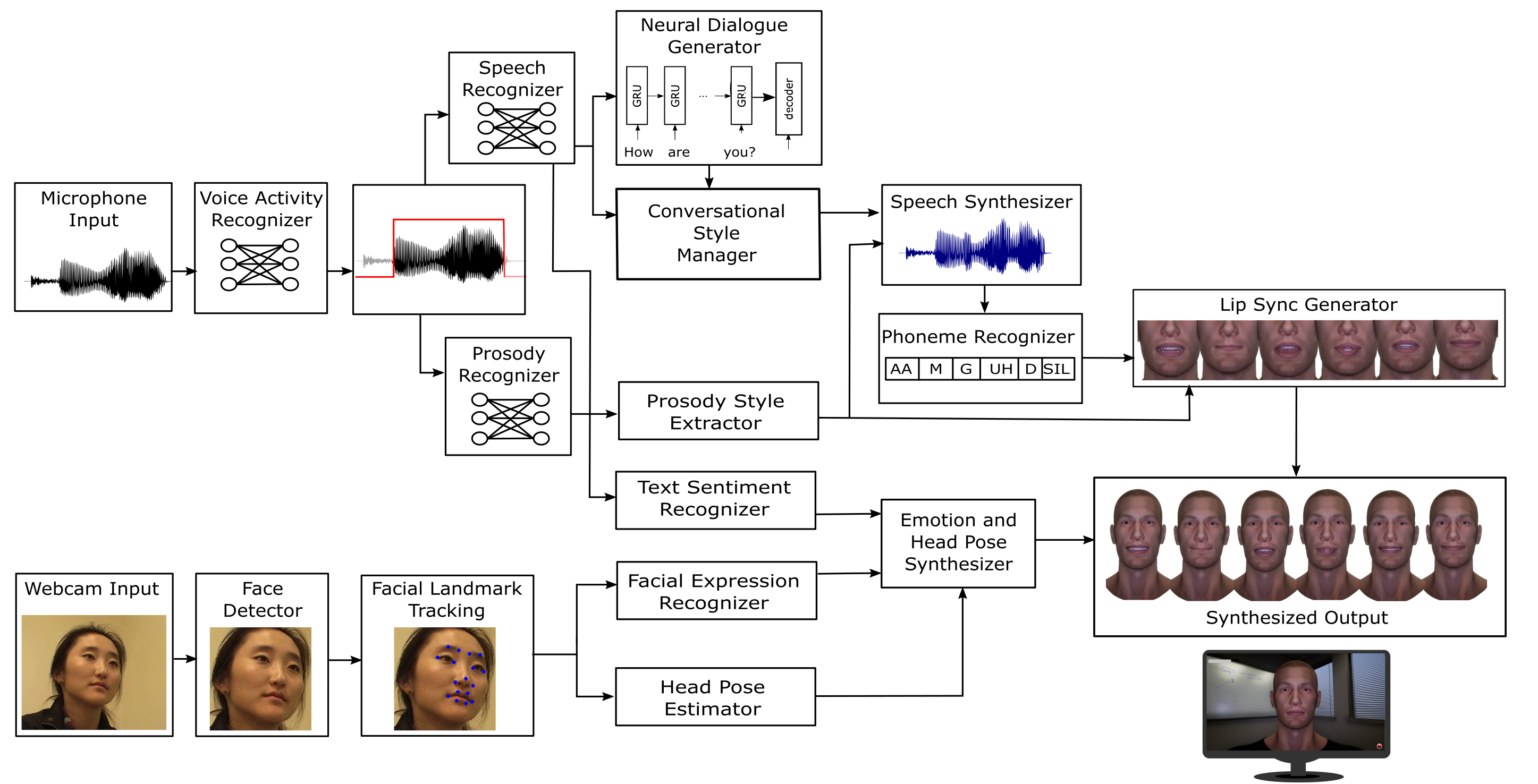}}
\caption{Our embodied conversational agent architecture. The voice-based agent architecture was similar but did not feature the virtual embodied avatar.}
\label{fig:architecture}
\vskip -0.2in
\end{figure}

\section{Challenges}
In this section, we discuss some of the important challenges in building conversational agents.

\subsubsection{Real-time performance} 
In the system diagram in Figure~\ref{fig:architecture} each of the recognizer and generator components was built using state-of-the-art deep neural networks, and as such, the computational cost was considerable. For example, even if each of the components in the dialogue pipeline (speech recognition, dialogue generation, and speech synthesis) only took approximately 700ms to execute (which is quite common for a deep network model), it would still take the agent a minimum of 2 seconds to respond. For modeling a HC conversation style, which typically would involve more frequent and longer pauses, this is not too much of a problem. However, in many human conversational interactions, and certainly when modeling the alternative, HI conversational style, it is desirable to deliver a response in less time than this. Modeling fast responses is very difficult if using the most accurate, state of the art, models.

\subsection{Voice-based agent}

\subsubsection{Subtlety of conversational style matching} One of the main challenges for an agent that matches the conversational style of an interlocutor is determining whether changes in the agent's behavior should be magnified to make them obvious to the conversational partner.  Calibrating these behaviors is challenging, and this is true for both their magnitude and temporal dynamics.  If style matching occurs too quickly, it might sound unnatural or forced. At the same time, matching too slowly could cause a confusing effect, where the agent responds too late to specific changes in speech (e.g., the user spoke excitedly about something, but moved on while the agent was only starting to adjust the user's pitch).

\subsubsection{Obtaining a baseline of conversational style} In order to match specific stylistic variables, we first need to obtain a baseline. For example, to match variance on loudness and pitch, we first need to obtain the averages of these. In general, participants interacted with the agent for a fairly extensive time, giving us enough observations to establish these baselines. However, how should the agent behave in the meantime, while these baselines are established? What is the optimal time window for establishing stylistic variables?  

\subsubsection{Conversational style can amplify negative effects in a conversation} In general, for our studies we hypothesized that conversational style matching would have a positive effect on how participants experienced the conversations with the agent. However, the agent matches any conversational style. So, it is possible that if a participant gets upset, the agent matches their angered speech patterns and as such the conversation could become much more negative. This might occur when a user gets frustrated with the agent (e.g., dissatisfied with answers or due to overlapping speech), in this case, directly matching conversation style could very well have an adverse effect on the conversation.

\subsubsection{Topic of discussions with a voice-based agent} People have gotten used to speaking with voice-based assistants like Siri, Alexa or Cortana due to their wide availability. However, most conversations with these types of assistants tend to be simple commands which need to be performed by the assistant. For our studies, we required participants to interact with the agent for an extended period of time. Therefore we designed the studies to involve having a social conversation with the agent. Since people are not used to having these types of conversations with an agent, we gave them some general scenarios to discuss. In spite of this, the participants mentioned experiencing this form of conversational interaction as a somewhat unnatural event, as it doesn't exist in shipping products today.

\subsubsection{Using a generative neural language model to generate responses} In order to perform conversational style matching, we wanted to provide proper responses to the user--however, we wanted to match participants on linguistic style (e.g., personal pronoun usage). As such, we used a generative neural language model approach to generate the answers. This model usually gives several similar responses to input and therefore allows us to select the proper response using conversational style matching. However, the downside to this model is that it tends to give very neutral responses and does not steer the conversation naturally. Because of this, the participants could run out of things to say very quickly. We, therefore, defined several scripted responses that would be triggered by the intent of the participant, using the Language Understanding and Intent Service (LUIS). These scripted responses would then generally return a question for the participant to answer. The downside to this, however, was that we could not change the scripted responses with the appropriate linguistic style matching.

\subsubsection{Misunderstanding participant input} Occasionally the speech recognition system would misunderstand the participant's input. This would cause the system to give a response that often would not make sense to the participant. 

\subsection{ECA}

\subsubsection{Realism} It is challenging to achieve realism with embodied characters without falling into the uncanny valley. We should render the ECA's face to display realism using appealing features consistently. For our ECA, we used healthy and natural skin tone, clear gender cues, natural and realistic proportions, and volumetric and healthy-colored lips, in order to avoid uncanny perceptions by the participants. 

\subsubsection{The conflict between speech generation and expressions} We needed to design the ECA to avoid verbal and visual conflicts like "the agent might be delivering a negative message while visually appearing happy." In our ECA design, we synthesized the expressions based on the text sentiment and synchronized the ECA's lips with the synthesized speech. One of the possible ways to avoid this conflict is to control the expressiveness of the ECA's upper face only (above the mouth) while lip syncing.

\subsubsection{ECA's visual response}  It's difficult to generate a perceptually valid ECA's visual response that is perceived as intended by the user during the conversation. In our current design, the ECA mimics the user's expressions while listening to the user. Mimicking of expressions limits the emotional expressiveness of agent and is not necessarily how humans respond to each other during conversation. Ideally, the agent's visual response should be empathetic and believable to build better rapport with the user. In our user study, two participants did not like the mimicking of their expressions and were a little uncomfortable during the interaction. We could overcome this limitation by generating the ECA's emotionally expressive visual response from a combination of multimodal input signals (audio, physiological, video) for a better understanding of the context and not just by mirroring the user's face.

\subsubsection{The ECA's expressiveness evaluation} We need standardized metrics to evaluate the ECA's expressive response better. To do that, we designed our questionnaire to measure the degree of believable nonverbal behaviour exhibited by the agent. Questions included: "To what degree did the agent reflect your emotions?", "To what degree did the agent exhibit believable nonverbal behavior?" and "To what degree did the agent emotionally connect with you?" on a scale of 1-7.


\section{User perceptions}
We describe how the interaction with a conversational agent affects the users' perception of the agent and their behavior during interaction.

\subsection{Voice-based agent}
Based on the post-study surveys with the voice-based agent (for both experimental and control condition), participants were fairly neutral on the agent. Using the Godspeed questionnaire~\cite{bartneck2009measurement} we measured participant's impressions of the agent and found that there were no significant differences based on the experimental condition. The first scale relating to Anthropomorphism, the mean rating of participants was a 2.2 out of 5 (higher scores being better), which shows that most participants did not consider the agent very human-like. The agent scored similarly on the scales of animacy (2.9) and perceived intelligence (2.7). On the likeability, the agent scored slightly higher, a 3.4. So although the agent was not considered all that human-like or alive, some participants did like the agent.


\subsection{ECA}
We analyzed the four Godspeed indices (anthropomorphism, animacy, likeability and perceived intelligence) as composite measures obtained through the survey ratings for each index. The animacy results indicated a significant difference in perception of the ECA for HI participants. The results also show that the participants perceived the agent to exhibit more convincing non-verbal behavior for the style matching condition (4.33 $\pm$ 1.59) compared to the controlled condition (2.21 $\pm$ 1.37) in a statistically significant way, ($t(27) = 3.83, p = 0.0006$). Overall, the results show that the expressions made the agent appear more animated and anthropomorphic for HI participants; however, for HC participants they did not. It is possible that, since the facial expressions of the agent were not entirely natural, the HC participants found them uncanny and the negative response to this was greater than the benefits of emotional expressions. Whereas, for the HI participants the opposite was true. 


\section{Design guidelines}
We define several design recommendations for designing a style matching agent using unconstrained dialogue models. The guidelines are based on our observations of participants interacting with the agent.

\begin{itemize}
\item {It would be wise to notify the participant of any significant system flaws, and not oversell the system's capabilities (e.g., delay in response).}

\item{In our work, the agent is designed to mimic the expressions of the user. It would be interesting to explore various other ways of generating appropriate expressions for the agent by understanding and learning from multi-modal input signals (audio, video, physiological).}
\item {As observed in our user studies, there was considerable overlap between the speech of the agent and the participant. Several fairly simple changes could deal with these types of issues in a system like this. One solution is allowing the agent to be interrupted (i.e., the agent stops talking when there is overlap). Another way of avoiding this type of overlap is to filter out stop words and interjections from the participant. Finally, the agent could provide feedback that s/he was "thinking," and not ready to reply yet.}
\item {Implementation of turn-taking and incorporation of pauses can improve the interaction with the user.}
\end{itemize}

%
\bibliographystyle{ACM-Reference-Format}
\bibliography{sample-base}

\end{document}